\documentclass[11pt]{article}
\usepackage{cite}
\usepackage{mathrsfs}
\usepackage{amssymb,amsfonts,amsmath}
\usepackage{enumerate}
\usepackage{indentfirst}
\usepackage{latexsym,bm}
\usepackage[noend]{algpseudocode}
\usepackage{algorithmicx,algorithm}
\usepackage{algorithm}
\usepackage{makeidx}
\usepackage{fancybox}
\usepackage{color}
\usepackage{multicol}
\usepackage[latin1]{inputenc}
\usepackage{graphicx}
\usepackage{pstricks,pst-node,pst-text,pst-3d}
\usepackage{tikz}
\newtheorem{thm}{Theorem}[section]

\newtheorem{pro}[thm]{Proposition}

\newenvironment{pf}{{\noindent \it \bf Proof:}}{{\hfill$\Box$}\\}

\begin{document}

\title{\bf Strong subgraph $k$-connectivity bounds}

\author{Yuefang Sun$^{1}$ and Gregory Gutin$^{2}$ \\
$^{1}$ Department of Mathematics,
Shaoxing University\\
Zhejiang 312000, P. R. China, yuefangsun2013@163.com\\
$^{2}$ Department of Computer Science\\
Royal Holloway, University of London\\
Egham, Surrey, TW20 0EX, UK, g.gutin@rhul.ac.uk}

\date{}
\maketitle

\begin{abstract}
 Let $D=(V,A)$ be a
digraph of order $n$, $S$ a subset of $V$ of size $k$ and
$2\le k\leq n$. Strong subgraphs $D_1, \dots , D_p$ containing $S$
are said to be internally disjoint if $V(D_i)\cap V(D_j)=S$ and $A(D_i)\cap
A(D_j)=\emptyset$ for all $1\le i<j\le p$.
Let $\kappa_S(D)$ be the maximum number of
internally disjoint strong digraphs containing $S$ in $D$.
The strong subgraph $k$-connectivity is defined as
$$\kappa_k(D)=\min\{\kappa_S(D)\mid S\subseteq V, |S|=k\}.$$
A digraph $D=(V, A)$ is called minimally strong subgraph
$(k,\ell)$-connected if $\kappa_k(D)\geq \ell$ but for any arc
$e\in A$, $\kappa_k(D-e)\leq \ell-1$.
In this paper, we first give a sharp upper bound for the parameter
$\kappa_k(D)$ and then study the minimally strong subgraph
$(k,\ell)$-connected digraphs.
\end{abstract}


\section{Introduction}\label{sec:intro}

The generalized $k$-connectivity $\kappa_k(G)$ of a graph $G=(V,E)$
was introduced by Hager \cite{Hager} in 1985 ($2\le k\le |V|$).
For a graph $G=(V,E)$ and a set $S\subseteq V$ of at least two
vertices, an {\em $S$-Steiner tree} or, simply, an {\em $S$-tree}
 is a subgraph
$T$ of $G$ which is a tree with $S\subseteq V(T)$. Two $S$-trees
$T_1$ and $T_2$ are said to be {\em internally disjoint} if
$E(T_1)\cap E(T_2)=\emptyset$ and $V(T_1)\cap V(T_2)=S$. The {\em
generalized local connectivity} $\kappa_S(G)$ is the maximum number
of internally disjoint $S$-trees in $G$. For an integer $k$ with
$2\leq k\leq n$, the {\em generalized $k$-connectivity} is defined
as
$$\kappa_k(G)=\min\{\kappa_S(G)\mid S\subseteq V(G), |S|=k\}.$$
Observe that
$\kappa_2(G)=\kappa(G)$.
If $G$ is disconnected and vertices of $S$ are placed in different connectivity components, we have  $\kappa_S(G)=0$.
Thus, $\kappa_k(G)=0$ for a disconnected graph $G$.
Generalized
connectivity of graphs has become an established area in graph
theory, see a recent monograph \cite{Li-Mao5} by Li and Mao on
generalized connectivity of undirected graphs.

To extend generalized $k$-connectivity to directed graphs, Sun, Gutin, Yeo and Zhang
\cite{Sun-Gutin-Yeo-Zhang} observed that
in the definition of $\kappa_S(G)$, one can replace ``an $S$-tree''
by ``a connected subgraph of $G$ containing $S$.'' Therefore,  Sun et al. \cite{Sun-Gutin-Yeo-Zhang}
defined {\em strong subgraph $k$-connectivity} by replacing
``connected'' with ``strongly connected'' (or, simply, ``strong'')
as follows. Let $D=(V,A)$ be a
digraph of order $n$, $S$ a subset of $V$ of size $k$ and
$2\le k\leq n$. Strong subgraphs $D_1, \dots , D_p$ containing $S$
are said to be {\em
internally disjoint} if $V(D_i)\cap V(D_j)=S$ and $A(D_i)\cap
A(D_j)=\emptyset$ for all $1\le i<j\le p$.
Let $\kappa_S(D)$ be the maximum number of
internally disjoint strong digraphs containing $S$ in $D$. The {\em
strong subgraph $k$-connectivity} is defined as
$$\kappa_k(D)=\min\{\kappa_S(D)\mid S\subseteq V, |S|=k\}.$$
By definition, $\kappa_2(D)=0$ if $D$ is not strong.

Despite the definition of strong subgraph $k$-connectivity being similar to that of generalized $k$-connectivity,
the former is somewhat more complicated than the latter. Let us first consider a simple reason for our claim above.
For a graph $G$, let $\overleftrightarrow{G}$ denote the digraph obtained from $G$ by replacing every edge $xy$ with two arcs $xy$ and $yx$.
While minimal connected spanning subgraphs of undirected graphs are all trees, even  a simple digraph $\overleftrightarrow{C_n}$  has two types of such strong subgraphs: a directed cycle and $\overleftrightarrow{P_n}$. A less trivial reason is given in the next paragraph. 

The main aim of \cite{Sun-Gutin-Yeo-Zhang} was to study complexity
of computing $\kappa_k(D)$ for an arbitrary digraph $D$, for a
semicomplete digraph $D$, and for a symmetric digraph $D$. In
particular, Sun et al. proved that for all fixed integers $k\ge 2$
and $\ell\ge 2$ it is NP-complete to decide whether $\kappa_S(D)\ge
\ell$ for an arbitrary digraph $D$ and a vertex set $S$ of $D$ of
size $k$. Since deciding the same problem for generalized
$k$-connectivity of undirected graphs is polynomial
time solvable \cite{Li-Li}, it is clear that computing strong
subgraph $k$-connectivity is somewhat harder than computing
generalized $k$-connectivity.

We will postpone discussion of further results from \cite{Sun-Gutin-Yeo-Zhang}
until Subsection \ref{sec:compres} and now overview new results obtained in this paper.
First, we improve the following tight bound used in \cite{Sun-Gutin-Yeo-Zhang}
\begin{equation}\label{eq:1}
\kappa_k(D)\le \min\{\delta^-(D),\delta^+(D)\}
\end{equation}
 for a digraph $D$, where $\delta^-(D)$ and
$\delta^+(D)$ are the minimum in-degree and out-degree of $D$, respectively.
We will show a new sharp bound $\kappa_k(D)\le \kappa(D)$, where  $\kappa(D)$ is the strong connectivity of $D$.
Note that $\kappa(D)\le  \min\{\delta^-(D),\delta^+(D)\}.$ Interestingly, for undirected graphs $G$, $\kappa_k(G)\le \kappa(G)$ holds only for $k\le 6$ \cite{Li,Li-Mao}.

In what follows, $n$ will denote the number of vertices of the digraph under consideration.

A digraph $D=(V(D), A(D))$ is called {\em minimally strong subgraph
$(k,\ell)$-connected} if $\kappa_k(D)\geq \ell$ but for any arc
$e\in A(D)$, $\kappa_k(D-e)\leq \ell-1$.
Let $\mathfrak{F}(n,k,\ell)$ be the set of all minimally strong
subgraph $(k,\ell)$-connected digraphs with order $n$. We define
$$F(n,k,\ell)=\max\{|A(D)| \mid D\in \mathfrak{F}(n,k,\ell)\}$$ and
$$f(n,k,\ell)=\min\{|A(D)| \mid D\in \mathfrak{F}(n,k,\ell)\}.$$ We
further define $$Ex(n,k,\ell)=\{D\mid D\in \mathfrak{F}(n,k,\ell),
|A(D)|=F(n,k,\ell)\}$$ and $$ex(n,k,\ell)=\{D\mid D\in
\mathfrak{F}(n,k,\ell), |A(D)|=f(n,k,\ell)\}.$$

Using the Hamilton cycle decomposition theorem of Tillson
\cite{Tillson}, Theorem \ref{thm01}, it is not hard to see
$f(n,k,n-1)=F(n,k,n-1)=n(n-1)$ and that the only extremal digraph is
the complete digraph on $n$ vertices. However, computing
$f(n,k,n-2)$ and $F(n,k,n-2)$ appears to be harder. In Theorem
\ref{thme}, we characterize minimally
strong subgraph $(2,n-2)$-connected digraphs.
The characterization implies that
$f(n,2,n-2)=n(n-1)-2\lfloor
n/2\rfloor$, $F(n,2,n-2)=n(n-1)-3$. We will also prove the
lower bound $f(n,k,\ell)\ge n\ell$ and describe some cases when
$f(n,k,\ell)= n\ell$. Finally, we will show that
$F(n,n,\ell)\leq 2\ell(n-1)$ and $F(n,k,1)=2(n-1).$
We leave it as an open problem to obtain a sharp upper bound on
$F(n,k,\ell)$ for every $k\ge 2$ and $\ell \ge 2$.

\subsection{Algorithms and Complexity Results} \label{sec:compres}

Let $k \ge 2$ and $\ell\ge 2$ be fixed integers. By reduction from
the {\sc Directed 2-Linkage} problem, Sun et al.
\cite{Sun-Gutin-Yeo-Zhang} proved that deciding whether
$\kappa_S(D)\ge \ell$ is NP-complete for a $k$-subset $S$ of $V(D)$.
Thomassen \cite{Thom} showed that for every positive integer $p$
there are digraphs which are strongly $p$-connected, but which
contain a pair of vertices not belonging to the same cycle. This
implies that for every positive integer $p$ there are digraphs $D$
such that $\kappa_2(D)=1$ \cite{Sun-Gutin-Yeo-Zhang}.

The above negative results motivate studying strong subgraph
$k$-connectivity for special classes of digraphs. In
\cite{Sun-Gutin-Yeo-Zhang}, Sun et al. showed that
the problem of deciding whether $\kappa_k(D)\ge \ell$ for every
semicomplete digraphs is polynomial-time solvable for fixed $k$ and
$\ell$. The main tool used in their proof is a recent {\sc Directed
$k$-Linkage} theorem 
of Chudnovsky, Scott and Seymour
\cite{Chud-Scott-Seymour}.

A digraph $D$ is {\em symmetric} if for every arc
$xy$ of $D$, $D$ also contains the arc $yx$. In other words, a
symmetric digraph $D$ can be obtained from its underlying undirected
graph $G$ by replacing each edge of $G$ with the corresponding arcs
of both directions, that is, $D=\overleftrightarrow{G}.$ Sun et al.
\cite{Sun-Gutin-Yeo-Zhang} showed that for any connected graph $G$,
the parameter $\kappa_2(\overleftrightarrow{G})$ can be computed in
polynomial time. This result is best possible in the following
sense, unless P$=$NP. Let $D$ be a symmetric digraph and $k\geq 3$ a
fixed integer. Then it is NP-complete to decide whether
$\kappa_S(D)\geq \ell$ for $S\subseteq V(D)$ with $|S|=k$
\cite{Sun-Gutin-Yeo-Zhang}.



\section{New sharp upper bound of $\kappa_k(D)$}\label{sec:smallk}

To prove a new bound on $\kappa_k(D)$ in Theorem \ref{thmb}, we will use the following proposition of Sun et al. \cite{Sun-Gutin-Yeo-Zhang}.

\begin{pro}\label{thm03}
Let $2\leq k\leq n$. For a strong digraph $D$ of order $n$, we have
$$1\leq \kappa_k(D)\leq n-1.$$ Moreover, both bounds are sharp, and
the upper bound holds if and only if $D\cong
\overleftrightarrow{K}_n$, $2\leq k\leq n$ and $k\not\in \{4,6\}$.
\end{pro}

\begin{thm}\label{thmb}
For $k\in \{2,\dots ,n\}$ and $n\ge \kappa(D)+k,$ we have
$$\kappa_k(D)\leq \kappa(D).$$ Moreover, the bound is sharp.
\end{thm}
\begin{pf}
For $k=2$, assume that $\kappa(D)=\kappa(x,y)$ for some
$\{x,y\}\subseteq V(D)$. It follows from the strong subgraph connectivity definition that
$\kappa_{\{x,y\}}(D)\leq \kappa(x,y)$, so $\kappa_2(D)\leq
\kappa_{\{x,y\}}(D)\leq \kappa(x,y)= \kappa(D).$

We now consider the case of $k\ge 3$. If $\kappa(D)=n-1$, then we
have $\kappa_k(D)\leq n-1=\kappa(D)$ by Proposition
\ref{thm03}. If $\kappa(D)=n-2$, then there two vertices, say $u$
and $v$, such that $uv\not\in A(D)$. So we have $\kappa_k(D)\leq
n-2=\kappa(D)$ by Proposition \ref{thm03}. If
$1\leq \kappa(D)\leq n-3$, then there exists a $\kappa(D)$-vertex
cut, say $Q$, for two vertices $u,v$ in $D$ such that there is no
$u-v$ path in $D-Q$. Let $S=\{u,v\}\cup S'$ where $S'\subseteq
V(D)\setminus (Q\cup \{u,v\})$ and $|S'|=k-2$. Since $u$ and $v$ are
in different strong components of $D-Q$, any strong subgraph
containing $S$ in $D$ must contain a vertex in $Q$. By the
definition of $\kappa_S(D)$ and $\kappa_k(D)$, we have
$\kappa_k(D)\leq \kappa_S(D)\leq |Q|=\kappa(D)$.

For the sharpness of the bound, 
consider the following digraph $D$. Let $D$ be a symmetric digraph
whose underlying undirected graph is $K_{k}\bigvee
\overline{K}_{n-k}$~($n\geq 3k$), i.e. the graph obtained from
disjoint graphs $K_{k}$ and $\overline{K}_{n-k}$ by adding all edges
between the vertices in $K_{k}$ and $\overline{K}_{n-k}$. 

Let $V(D)=W\cup U$, where $W=V(K_k)=\{w_i\mid 1\leq i\leq k\}$ and
$U=V(\overline{K}_{n-k})=\{u_j\mid 1\leq j\leq n-k\}$. Let $S$ be any $k$-subset of
vertices of $V(D)$ such that $|S\cap U|=s$ ($s\leq k$) and $|S\cap
W|=k-s$. Without loss of generality, let $w_i\in S$ for $1\leq i\leq
k-s$ and $u_j\in S$ for $1\leq j\leq s$. For $1\leq i\leq k-s$, let
$D_i$ be the symmetric subgraph of $D$ whose underlying undirected
graph is the tree $T_i$ with edge set $$\{w_iu_1, w_iu_2, \dots , w_iu_s,
u_{k+i}w_1, u_{k+i}w_2, \dots , u_{k+i}w_{k-s}\}.$$ For
$k-s+1\leq j\leq k$, let $D_j$ be the symmetric subgraph of $D$
whose underlying undirected graph is the tree $T_j$ with edge set $$\{w_ju_1,
w_ju_2, \dots , w_ju_s, w_jw_1, w_jw_2, \dots ,
w_jw_{k-s}\}.$$ Observe that $\{D_i\mid 1\leq i\leq
k-s\}\cup \{D_j\mid k-s+1\leq j\leq k\}$ is a set of $k$ internally
disjoint strong subgraph containing $S$, so $\kappa_S(D)\geq k$, and
then $\kappa_k(D)\geq k$. Combining this with the bound that
$\kappa_k(D)\leq \kappa(D)$ and the fact that $\kappa(D)\leq
\min\{\delta^+(D), \delta^-(D)\}=k$, we can get $\kappa_k(D)=
\kappa(D)=k$.
\end{pf}


\section{Minimally strong subgraph $(k,\ell)$-connected digraphs}\label{sec:minimally}

Below we will use the following Hamilton cycle decomposition theorem of
Tillson.
\begin{thm}\cite{Tillson}\label{thm01}
The arcs of $\overleftrightarrow{K}_n$ can be decomposed into
Hamiltonian cycles if and only if $n\neq 4,6$.
\end{thm}



The following observation will be used in the sequel.

\begin{pro}\label{thm02}\cite{Sun-Gutin-Yeo-Zhang}
If $D'$ is a strong spanning digraph of a strong digraph $D$, then
$\kappa_k(D')\leq \kappa_k(D)$.
\end{pro}

By the definition of a minimally strong subgraph $(k,\ell)$-connected
digraph, we can get the following observation.

\begin{pro}\label{lem1}
A digraph $D$ is minimally strong subgraph $(k,\ell)$-connected if
and only if $\kappa_k(D)= \ell$ and $\kappa_k(D-e)= \ell-1$ for any
arc $e\in A(D)$.
\end{pro}
\begin{pf} The direction ``if" is clear by definition, and we only
need to prove the direction ``only if". Let $D$ be a minimally
strong subgraph $(k,\ell)$-connected digraph. By definition, we have
$\kappa_k(D)\geq \ell$ and $\kappa_k(D-e)\leq \ell-1$ for any arc
$e\in A(D)$. Then for any set $S \subseteq V(D)$ with $|S|=k$, there
is a set $\mathcal{D}$ of $\ell$ internally disjoint strong subgraphs containing
$S$. As $e$ must belong to one and only one element of $\mathcal{D}$, we are done.
\end{pf}


A digraph $D$ is {\em minimally strong} if $D$ is strong but $D-e$ is not for every arc $e$ of $D$.

\begin{pro}\label{thmc}
The following assertions hold:\\
$(i)$~A digraph $D$ is minimally strong subgraph $(k,1)$-connected
if and only if $D$ is minimally strong digraph;\\
$(ii)$~For $k\neq 4,6$, a digraph $D$ is minimally strong subgraph
$(k,n-1)$-connected if and only if $D\cong
\overleftrightarrow{K}_n$.
\end{pro}
\begin{pf}
To prove (i), it  suffices to show that a digraph $D$ is strong if and only if $\kappa_k(D)\ge 1.$ If $D$ is strong, then for every vertex set $S$ of size $k,$ $D$ has a strong subgraph containing $S$. If $\kappa_k(D)\ge 1$, for each vertex set $S$ of size $k$ construct $D_S,$ a strong subgraph of $D$ containing $S.$ The union of all $D_k$ is a strong subgraph of $D$ as there are sets $S_1, S_2, \dots , S_p$ such that the union of $S_1, S_2, \dots , S_p$ is $V(D)$ and for each $i\in [p-1],$ $D_{S_i}$ and $D_{S_{i+1}}$ share a common vertex.

Part (ii) follows from Proposition \ref{thm03}.
\end{pf}

The following result characterizes minimally
strong subgraph $(2,n-2)$-connected digraphs.

\begin{thm}\label{thme}
A digraph $D$ is minimally strong subgraph $(2,n-2)$-connected if and only if $D$ is a digraph obtained from the complete digraph
$\overleftrightarrow{K}_n$ by deleting an arc set M such that $\overleftrightarrow{K}_n[M]$ is a 3-cycle
or a union of $\lfloor n/2\rfloor$ vertex-disjoint 2-cycles. In particular,
we have $f(n,2,n-2)=n(n-1)-2\lfloor n/2\rfloor$, $F(n,2,n-2)=n(n-1)-3$.
\end{thm}
\begin{pf}
Let $D\cong \overleftrightarrow{K}_n-M$ be a digraph obtained from
the complete digraph $\overleftrightarrow{K}_n$ by deleting an arc
set $M$. Let $V(D)=\{u_i\mid 1\leq i\leq n\}$.

Firstly, we will
consider the case that $\overleftrightarrow{K}_n[M]$ is a 3-cycle
$u_1u_2u_3u_1$. We now prove that $\kappa_2(D)=n-2$. By
(\ref{eq:1}), we have $\kappa_2(D)\leq \min\{\delta^+(D),
\delta^-(D)\}=n-2$. Let $S=\{u,v\} \subseteq V(D)$; we just consider
the case that $u=u_1,v=u_2$ since the other cases are similar. Let
$D_1$ be a subdigraph of $D$ with $V(D_1)=\{u_1,u_2,u_3\}$ and
$A(D_1)=\{u_1u_3, u_3u_2, u_2u_1\}$; for $2\leq i\leq n-2$, let
$D_i$ be a subdigraph of $D$ with $V(D_i)=\{u_1,u_2,u_{i+2}\}$ and
$A(D_i)=\{u_1u_{i+2}, u_2u_{i+2}, u_{i+2}u_1, u_{i+2}u_2\}$.
Clearly, $\{D_i\mid 1\leq i\leq n-2\}$ is a set of $n-2$ internally
disjoint strong subgraphs containing $S$, so $\kappa_S(D)\geq n-2$
and $\kappa_2(D)\geq n-2$. Hence, $\kappa_2(D)= n-2$.

For any $e\in A(D)$, without loss of generality, one of the two digraphs in Figure \ref{figure1}
is a subgraph of $\overleftrightarrow{K}_n[M\cup \{e\}]$,
so if the following claim holds, then we must have
$\kappa_2(D-e)\leq \kappa_2(D') \leq n-3$ by Proposition
\ref{thm02}, and so $D$ is minimally strong subgraph
$(2,n-2)$-connected. Now it suffices to prove the following claim.

\begin{figure}[!hbpt]
\begin{center}
\includegraphics[scale=0.80]{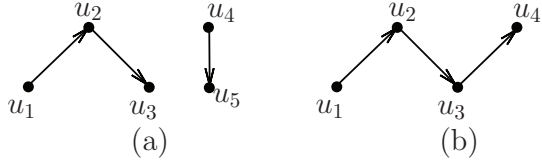}
\end{center}
\caption{Two graphs for Claim 1.}\label{figure1}
\end{figure}

\noindent  \textbf {Claim 1.} If $\overleftrightarrow{K}_n[M']$ is isomorphic
to one of two graphs in Figure \ref{figure1}, then $\kappa_2(D')\leq
n-3$, where $D'=\overleftrightarrow{K}_n-M'$.

\noindent  {\it Proof of Claim 1.} We first show that $\kappa_2(D')\leq n-3$ if
$M'$ is the digraph of Figure \ref{figure1} $(a)$. Let
$S=\{u_2, u_4\}$; we will prove that $\kappa_S(D')\leq n-3$, and
then we are done. Suppose that $\kappa_S(D')\geq n-2$, then there
exists a set of $n-2$ internally disjoint strong subgraphs
containing $S$, say $\{D_i\mid 1\leq i\leq n-2\}$. If both of the
two arcs $u_2u_4$ and $u_4u_2$ belong to the same $D_i$, say $D_1$,
then for $2\leq i\leq n-2$, each $D_i$ contains at least one vertex
and at most two vertices of $\{u_i\mid 1\leq i\leq n, i\neq 2,4\}$.
Furthermore, there is at most one $D_i$, say $D_2$, contains
(exactly) two vertices of $\{u_i\mid 1\leq i\leq n, i\neq 2,4\}$. We
just consider the case that $u_1,u_3\in V(D_2)$ since the other
cases are similar. In this case, we must have that each vertex of
$\{u_i\mid 5\leq i\leq n\}$ belongs to exactly one digraph from
$\{D_i\mid 3\leq i\leq n-2\}$ and vice versa.
However, this is impossible since the vertex set $\{u_2, u_4,
u_5\}$ cannot induce a strong subgraph of $D'$ containing $S$, a
contradiction.

So we now assume that each $D_i$ contains at most one
of $u_2u_4$ and $u_4u_2$. Without loss of generality, we may assume that
$u_2u_4\in A(D_1)$ and $u_4u_2\in A(D_2)$. In this case, we must
have that each vertex of $\{u_i\mid 1\leq i\leq n, i\neq 2,4\}$
belongs to exactly one digraph from $\{D_i\mid 1\leq i\leq n-2\}$ and vice versa.
However, this is
also impossible since the vertex set $\{u_2, u_4, u_5\}$ cannot
induce a strong subgraph of $D'$ containing $S$, a contradiction.

Hence, we
have $\kappa_2(D')\leq n-3$ in this case. For the case that $M'$ is
 the digraph of Figure \ref{figure1}  $(b)$, we can choose
$S=\{u_2, u_3\}$ and prove that $\kappa_S(D')\leq n-3$ with a
similar argument, and so $\kappa_2(D')\leq n-3$ in this case. This
completes the proof of the claim.

\vspace{2mm}
Secondly, we consider the case that  $\overleftrightarrow{K}_n[M]$
is a union of $\lfloor n/2\rfloor$ vertex-disjoint 2-cycles. Without
loss of generality, we may assume that $M=\{u_{2i-1}u_{2i}, u_{2i}u_{2i-1}\mid 1\leq
i\leq \lfloor n/2\rfloor\}$. We just consider the case that
$S=\{u_1, u_3\}$ since the other cases are similar. In this case,
let $D_1$ be the subgraph of $D$ with $V(D_1)=\{u_1, u_3\}$ and
$A(D_1)=\{u_1u_3, u_3u_1\}$; let $D_2$ be the subgraph of $D$ with
$V(D_2)=\{u_1,u_2,u_3,u_4\}$ and $A(D_2)=\{u_1u_4, u_4u_1, u_2u_4,
u_4u_2, u_2u_3, u_3u_2\}$; for $3\leq i\leq n-2$, let $D_i$ be the
subgraph of $D$ with $V(D_i)=\{u_1,u_2,u_{i+2}\}$ and
$A(D_i)=\{u_1u_{i+2}, u_3u_{i+2}, u_{i+2}u_1, u_{i+2}u_3\}$.
Clearly, $\{D_i\mid 1\leq i\leq n-2\}$ is a set of $n-2$ internally
disjoint strong subgraphs containing $S$, so $\kappa_S(D)\geq n-2$
and then $\kappa_2(D)\geq n-2$. By (\ref{eq:1}), we have
$\kappa_2(D)\leq \min\{\delta^+(D), \delta^-(D)\}=n-2$. Hence,
$\kappa_2(D)= n-2$. Let $e\in A(D)$; clearly $e$ must be incident with
at least one vertex of $\{u_i\mid 1\leq i\leq 2\lfloor
n/2\rfloor\}$. Then we have that $\kappa_2(D-e)\leq
\min\{\delta^+(D-e), \delta^-(D-e)\}=n-3$ by  (\ref{eq:1}).
Hence, $D$ is minimally strong subgraph $(2,n-2)$-connected.

\vspace{2mm} Now let $D$ be minimally strong subgraph
$(2,n-2)$-connected. By Proposition~\ref{thm03}, we
have that $D\not \cong \overleftrightarrow{K}_n$, that is, $D$ can
be obtained from a complete digraph $\overleftrightarrow{K}_n$ by
deleting a nonempty arc set $M$. To end our argument, we need the
following three claims. Let us start from a simple yet useful
observation.

\begin{pro}\label{pro:HT}
No pair of arcs in $M$ has
a common head or tail.
\end{pro}
{\it Proof of Proposition \ref{pro:HT}.}
By (\ref{eq:1}) no pair of arcs in $M$ has
a common head or tail, as otherwise we would have $\kappa_2(D)\le n-3$.

\vspace{3mm}

\noindent \textbf {Claim 2.} $|M|\geq 3$.

\noindent {\it Proof of Claim 2.} Let $|M|\le 2$. We may assume that $|M|=2$
as the case of $|M|=1$ can be considered in a similar and simpler way.

Let the arcs of $M$ have no common vertices; without loss of generality, $M=\{u_1u_2,u_3u_4\}$. Then $\kappa_2(D-u_2u_1)=n-2$
as  $D-u_2u_1$ is a supergraph of $\overleftrightarrow{K}_n$ without a union of $\lfloor n/2\rfloor$
vertex-disjoint 2-cycles including the cycles $u_1u_2u_1$ and $u_3u_4u_3$. Thus,  $D$ is not minimally strong
subgraph $(2,n-2)$-connected. Let the arcs of $M$ have no common vertex. By Proposition \ref{pro:HT}, without loss of generality, $M=\{u_1u_2,u_2u_3\}$.
Then $\kappa_2(D-u_3u_1)=n-2$ as we showed in the beginning of the proof of this theorem. Thus,  $D$ is not minimally strong
subgraph $(2,n-2)$-connected.
Now let the arcs of $M$ have the same vertices, i.e., without loss
of generality, $M=\{u_1u_2,u_2u_1\}$. As above, $\kappa_2(D-u_2u_1)=n-2$ and $D$ is not minimally strong
subgraph $(2,n-2)$-connected.

\vspace{3mm}

\noindent  \textbf {Claim 3.} If $|M|= 3$, then $\overleftrightarrow{K}_n[M]$
is a 3-cycle.

\noindent  {\it Proof of Claim 3.} Suppose that $D$ is minimally strong subgraph
$(2,n-2)$-connected, but $\overleftrightarrow{K}_n[M]$
is not a 3-cycle.
By Proposition \ref{pro:HT}, no pair of arcs in $M$ has
a common head or tail.
Thus, $\overleftrightarrow{K}_n[M]$ must be isomorphic to one of graphs in
Figures \ref{figure1} and \ref{figure3}. If
$\overleftrightarrow{K}_n[M]$ is isomorphic to one of graphs in
Figure \ref{figure1}, then $\kappa_2(D)\leq n-3$ by Claim 1 and so
$D$ is not minimally strong subgraph $(2,n-2)$-connected, a contradiction.
For an arc set $M_0$ such that
$\overleftrightarrow{K}_n[M_0]$ is a union of $\lfloor n/2\rfloor$
vertex-disjoint 2-cycles, by the argument before, we know that
$\overleftrightarrow{K}_n-M_0$ is minimally strong subgraph
$(2,n-2)$-connected. For the case that $\overleftrightarrow{K}_n[M]$
is isomorphic to $(a)$ or $(b)$ in Figure \ref{figure3}, we have
that $\overleftrightarrow{K}_n-M_0$ is a proper subdigraph of
$\overleftrightarrow{K}_n-M$, so $D=\overleftrightarrow{K}_n-M$ must
not be minimally strong subgraph $(2,n-2)$-connected, this also
produces a contradiction. Hence, the claim holds.

\begin{figure}[!hbpt]
\begin{center}
\includegraphics[scale=0.80]{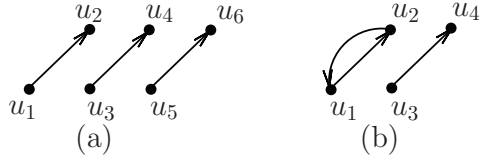}
\end{center}
\caption{Two graphs for Claim 3.}\label{figure3}
\end{figure}

\vspace{3mm}

\noindent \textbf {Claim 4.} If $|M|> 3$, then $\overleftrightarrow{K}_n[M]$
is a union of $\lfloor n/2\rfloor$ vertex-disjoint 2-cycles.

\noindent {\it Proof of Claim 4.} Suppose that $D$ is minimally strong subgraph
$(2,n-2)$-connected, but $\overleftrightarrow{K}_n[M]$
is not a union of $\lfloor n/2\rfloor$ vertex-disjoint 2-cycles.

By Claim 1 and Proposition \ref{thm02}, we
have that $\overleftrightarrow{K}_n[M]$ does not contain graphs in Figure
\ref{figure1} as a subgraph. Then $\overleftrightarrow{K}_n[M]$
does not contain a path of length
at least three. Hence, the underlying undirected graph of $M$ has at least
two connectivity components. By the fact that if $M$ is a
3-cycle, then $\overleftrightarrow{K}_n-M$ is minimally
strong subgraph $(2,n-2)$-connected, we conclude that
$\overleftrightarrow{K}_n[M]$ does not contain a cycle of length
three. By Claim 1, $\overleftrightarrow{K}_n[M]$ does not
contain a path of length two. By Proposition \ref{pro:HT}, no pair of arcs in $M$ has
a common head or tail. Hence, each connectivity component of
$\overleftrightarrow{K}_n[M]$ must be a 2-cycle or an arc. Since $D$ is
minimally strong subgraph $(2,n-2)$-connected, no connectivity component of $\overleftrightarrow{K}_n[M]$
is an arc. We have arrived at a contradiction, proving Claim 4.

Hence, if a digraph $D$ is minimally strong subgraph
$(2,n-2)$-connected, then $D\cong \overleftrightarrow{K}_n-M$, where
$\overleftrightarrow{K}_n[M]$ is a cycle of order three or a union
of $\lfloor n/2\rfloor$ vertex-disjoint 2-cycles.

Now the claimed values of $F(n,2,n-2)$ and $f(n,2,n-2)$ can easily be verified.
\end{pf}

Note that Theorem \ref{thme} implies that
$Ex(n,2,n-2)=\{\overleftrightarrow{K_n}-M\}$ where $M$ is an arc set
such that $\overleftrightarrow{K}_n[M]$ is a directed 3-cycle, and
$ex(n,2,n-1)=\{\overleftrightarrow{K_n}-M\}$ where $M$ is an arc set
such that $\overleftrightarrow{K}_n[M]$ is a union of $\lfloor
n/2\rfloor$ vertex-disjoint directed 2-cycles.

The following result concerns a sharp lower bound for the parameter
$f(n,k,\ell)$.

\begin{thm}\label{thma} For $2\leq k\leq n$, we have
$$f(n,k,\ell)\geq n\ell.$$
Moreover, the following assertions hold:\\
$(i)~$ If $\ell=1$, then $f(n,k,\ell)=n$; $(ii)~$ If $2\leq \ell\leq
n-1$, then $f(n,n,\ell)=n\ell$ for $k=n\not\in \{4,6\}$; (iii) If
$n$ is even and $\ell = n-2$, then
$f(n,2,\ell)=n\ell.$
\end{thm}
\begin{pf}
By  (\ref{eq:1}), for all digraphs $D$ and $k \geq 2$ we have
$\kappa_k(D) \leq \delta^+(D)$ and $\kappa_k(D) \leq \delta^-(D)$.
Hence for each $D$ with $\kappa_k(D)=\ell$, we have that
$\delta^+(D), \delta^-(D)\geq \ell$, so $|A(D)|\geq n\ell$ and then
$f(n,k,\ell)\geq n\ell.$

For the case that $\ell=1$, let $D$ be a dicycle
$\overrightarrow{C_n}$. Clearly, $D$ is minimally strong subgraph
$(k,1)$-connected, and we know $|A(D)|=n$, so $f(n,k,1)= n$.

For the case that $k=n \not\in \{4,6\}$ and $2\leq \ell\leq n-1$,
let $D\cong \overleftrightarrow{K_n}$. By Theorem \ref{thm01}, $D$
can be decomposed into $n-1$ Hamiltonian cycles $H_i~(1\leq i\leq
n-1)$. Let $D_{\ell}$ be the spanning subdigraph of $D$ with arc
sets $A(D_{\ell})=\bigcup_{1\leq i\leq \ell}{A(H_i)}$. Clearly, we
have $\kappa_n(D_{\ell})\geq \ell$ for $2\leq \ell\leq n-1$.
Furthermore, by (\ref{eq:1}), we have $\kappa_n(D_{\ell})\leq
\ell$ since the in-degree and out-degree of each vertex in $D_{\ell}$
are both $\ell$. Hence, $\kappa_n(D_{\ell})= \ell$ for $2\leq
\ell\leq n-1$. For any $e\in A(D_{\ell})$, we have
$\delta^+(D_{\ell}-e)=\delta^-(D_{\ell}-e)=\ell-1$, so
$\kappa_n(D_{\ell}-e)\leq \ell-1$ by (\ref{eq:1}).
Thus, $D_{\ell}$ is minimally strong
subgraph $(n,\ell)$-connected. As $|A(D_{\ell})|=n\ell$, we have
$f(n,n,\ell)\leq n\ell$. From the lower bound that $f(n,k,\ell)\geq
n\ell$, we have $f(n,n,\ell)= n\ell$ for the case that $2\leq
\ell\leq n-1, n\not\in \{4,6\}$.

Part (iii) follows directly from Theorem
\ref{thme}.
\end{pf}




To prove two upper bounds on the number of arcs in a minimally strong subgraph $(k,\ell)$-connected digraph, we will use the following result, see e.g. \cite{Bang-Jensen-Gutin}.

\begin{thm}\label{2n-2-thm}
Every strong digraph $D$ on $n$ vertices has a strong spanning subgraph $H$ with at most $2n-2$ arcs and equality holds only if $H$ is a symmetric digraph whose underlying undirected graph is a tree.
\end{thm}

\begin{pro}
We have
$(i)$~$F(n,n,\ell)\le 2\ell(n-1)$;
$(ii)$~For every $k$ $(2\le k\le n)$, $F(n,k,1)=2(n-1)$ and $Ex(n,k,1)$ consists of
symmetric digraphs whose underlying undirected graphs are trees.
\end{pro}
\begin{pf}
$(i)$~Let $D=(V,A)$ be a minimally strong subgraph
$(n,\ell)$-connected digraph, and let $D_1,\dots ,D_{\ell}$ be
arc-disjoint strong spanning subgraphs of $D$. Since $D$ is
minimally strong subgraph $(n,\ell)$-connected and $D_1,\dots
,D_{\ell}$ are pairwise arc-disjoint,
$|A|=\sum_{i=1}^{\ell}|A(D_i)|.$ Thus, by Theorem \ref{2n-2-thm},
$|A|\le 2\ell(n-1).$

$(ii)$~In the proof of  Proposition \ref{thmc} we showed that a digraph $D$ is strong if and only if $\kappa_k(D)\ge 1.$ Now let $\kappa_k(D)\ge 1$ and a digraph $D$ has a minimal number of arcs. By Theorem \ref{2n-2-thm}, we have that $|A(D)|\le 2(n-1)$ and if $D \in Ex(n,k,1)$  then $|A(D)|=2(n-1)$ and $D$ is a symmetric digraph whose underlying undirected graph is a tree.
\end{pf}

\section{Discussion}
Perhaps, the most interesting result of this paper is the characterization of minimally
strong subgraph $(2,n-2)$-connected digraphs. As a simple consequence of the characterization, we can
determine the values of $f(n,2,n-2)$ and $F(n,2,n-2)$. It would be interesting to determine $f(n,k,n-2)$ and $F(n,k,n-2)$ for every value of $k\ge 3$.
(Obtaining characterizations of all $(k,n-2)$-connected digraphs for $k\ge 3$ seems a very difficult problem.)
It would also be interesting to find a sharp upper bound for $F(n,k,\ell)$ for all $k\ge 2$ and $\ell\ge 2$.




\vskip 1cm \noindent {\bf Acknowledgements.} Yuefang Sun was
supported by National Natural Science Foundation of China (No.
11401389). Gregory Gutin was partially supported by Royal Society
Wolfson Research Merit Award.


\begin{thebibliography} {99}

\bibitem{Bang-Jensen-Gutin}
J. Bang-Jensen and G. Gutin, Digraphs: Theory, Algorithms and Applications, 2nd Edition, Springer, London, 2009.






\bibitem{Chud-Scott-Seymour} M. Chudnovsky, A. Scott and P.D. Seymour. Disjoint paths in unions of tournaments. arXiv:1604.02317, April 2016.



\bibitem{Hager} M. Hager, Pendant tree-connectivity, J. Combin. Theory Ser. B 38, 1985, 179--189.


\bibitem{Li} S. Li, Some topics on generalized connectivity of graphs, PhD thesis, Nankai University, 2012.

\bibitem{Li-Li}
S. Li and X. Li, Note on the hardness of generalized connectivity,
J. Comb. Optim. 24(3), 2012, 389--396.


\bibitem{Li-Mao} X. Li and Y. Mao, A survey on the generalized connectivity of graphs, arXiv:1207.1838, v10, Aug 2015.

\bibitem{Li-Mao5} X. Li and Y. Mao, Generalized Connectivity of
Graphs, Springer, Switzerland, 2016.







\bibitem{Sun-Gutin-Yeo-Zhang}
Y. Sun, G. Gutin, A. Yeo, X. Zhang, Strong subgraph
$k$-connectivity, submitted.


\bibitem{Thom} C. Thomassen, Highly connected non-2-linked digraphs, Combinatorica 11(4) (1991) 393--395.

\bibitem{Tillson}
T.W. Tillson, A Hamiltonian decomposition of $K^*_{2m}$, $2m \geq
8$, J. Combin. Theory Ser. B 29(1), 1980, 68--74.

\end{thebibliography}
\end{document}